\documentclass[aps,prd,nofootinbib,twocolumn,superscriptaddress]{revtex4}
\usepackage{graphicx}
\usepackage{amsmath}

\newcommand{\ie}{{\it i.e., }}

\newcommand{\be}{\begin{equation}}
\newcommand{\ee}{\end{equation}}
\newcommand{\br}{\begin{eqnarray}}
\newcommand{\bea}{\begin{eqnarray}}
\newcommand{\eea}{\end{eqnarray}}
\newcommand{\er}{\end{eqnarray}}
\newcommand{\ba}{\begin{array}}
\newcommand{\ea}{\end{array}}
\newcommand{\bn}{\begin{enumerate}}
\newcommand{\en}{\end{enumerate}}
\newcommand{\bc}{\begin{center}}
\newcommand{\ec}{\end{center}}

\newcommand{\Eq}[1]{Eq.~(\ref{#1})}
\newcommand{\rfn}[1]{(\ref{#1})}
\newcommand{\Luv}{\Lambda_{\rm{UV}}}
\newcommand{\Lir}{\Lambda_{\rm{IR}}}
\newcommand{\vw}{v_{\textrm{weak}}}

\newcommand{\beq}{\begin{equation}}
\newcommand{\eeq}{\end{equation}}

\newcommand{\gsim}{\lower.7ex\hbox{$\;\stackrel{\textstyle>}{\sim}\;$}}
\newcommand{\lsim}{\lower.7ex\hbox{$\;\stackrel{\textstyle<}{\sim}\;$}}

\def\mysection#1{{\bf #1.} }

\begin{document}

\title{Physical Naturalness and Dynamical Breaking of Classical Scale Invariance }

\author{Matti Heikinheimo}
\affiliation{National Institute of Chemical Physics and Biophysics, R\"avala 10, 10143 Tallinn, Estonia}

\author{Antonio Racioppi}
\affiliation{National Institute of Chemical Physics and Biophysics, R\"avala 10, 10143 Tallinn, Estonia}

\author{Martti Raidal}
\affiliation{National Institute of Chemical Physics and Biophysics, R\"avala 10, 10143 Tallinn, Estonia}
\affiliation{Institute of Physics, University of Tartu, Estonia}

\author{Christian Spethmann}
\affiliation{National Institute of Chemical Physics and Biophysics, R\"avala 10, 10143 Tallinn, Estonia}

\author{Kimmo Tuominen}
\affiliation{Helsinki Institute of Physics, P.O.Box 64, FIN-00014, Helsinki University, Finland}

\begin{abstract}
We propose a model of a confining dark sector, dark technicolor, that communicates with the Standard Model through the Higgs portal. In this model electroweak symmetry breaking and dark matter share
a common origin, and the electroweak scale is generated dynamically.
Our motivation to suggest this model is the absense of evidence for new physics from recent LHC data.
Although the conclusion is far from certain at this point, this lack of evidence may suggest that no mechanism exists at the electroweak scale to stabilise the Higgs mass against radiative corrections from UV physics. The usual reaction to this puzzling situation is to conclude that the stabilising new physics is either hidden from us by accident, or that it appears at energies that are currently inaccessible, such that nature is indeed fine-tuned. 
In order to re-examine the arguments that have lead to this dichotomy, we review the concept of naturalness in effective field theories, discussing in particular the role of quadratic divergences in relation to different energy scales.
This leads us to suggest classical scale invariance as a guidline for model building, implying that explicit mass scales are absent in the underlying theory.

\end{abstract}

\maketitle

\section{Introduction}
The discovery~\cite{Aad:2012tfa,Chatrchyan:2012ufa} of a Higgs boson~\cite{Englert:1964et,Higgs:1964ia,Higgs:1964pj,Guralnik:1964eu} at the LHC
proves experimentally that scalars play a fundamental role in particle physics.
The global fits to the ATLAS and CMS Higgs data, about 25/fb per experiment, and to the Tevatron final Higgs results
constrain the Higgs boson couplings to agree with standard model (SM) predictions within remarkable ${\cal O}(20\%)$
precision~\cite{Giardino:2013bma,Ellis:2013lra,Djouadi:2013qya,Falkowski:2013dza,Alanne:2013dra}.
New physics beyond the SM in the Higgs sector can only be a small perturbation around the SM.
This result agrees with the ATLAS and CMS null results in searches for supersymmetry and for any other new physics (compositeness, extra dimensions,
extended gauge sectors etc)---and also with the experimental results from flavour physics and electroweak precision data---that all indicate no 
new physics beyond the SM. At the same time, the existence of cold dark matter (DM)~\cite{Ade:2013lta} and neutrino oscillations~\cite{Strumia:2006db} 
provides firm evidence for the existence of new physics beyond the SM.
In the light of the LHC, this coexistence becomes increasingly difficult to explain within the usual new physics scenarios that have been dominating 
the model building landscape. In particular, the question of naturalness, \ie why the electroweak (EW) scale exists and what stabilises it
against radiative corrections from high scale physics, remains unanswered.

The hierarchy problem is often simplified as a problem of 
quadratic divergences. This is a convenient shorthand, but misses the real physical content of the issue. 
We therefore start our discussion by discussing ``physical naturalness'', which is 
is independent of specific regularisation schemes.
Mixing of operators due to renormalisation, including mixing of scales, implies that large radiative corrections to the Higgs mass may occur. In perturbation theory, such mixing originates from physical particles with masses at the UV scale that are coupled to IR physics.
All quadratic divergences occurring in cutoff regularisation need to be related to such
physical mass scales. 

For a long time, an extension of the group of space-time symmetries in the form of broken supersymmetry has played a central role 
in the development of models for physics beyond the SM. Supersymmetry has several desirable features, the most important of which is 
perhaps the stabilisation of the Higgs mass and the EW scale against radiative corrections. However, realistic models of natural electroweak symmetry
breaking based on supersymmetry require the existence of superpartners of the SM particles at or below the TeV scale, none  of which have been experimentally observed. It is intriguing to contemplate  instead  another extension of space-time symmetries, namely classical 
conformal symmetry broken down to the Lorentz group by quantum corrections.

According to this paradigm the SM Higgs appears light due to a 
perturbation of the underlying classical scale invariance as proposed by Bardeen~\cite{Bardeen:1995kv}, which otherwise would prohibit any mass 
scales. The idea of classical scale invariance is not new. Already in 1973 Coleman and Weinberg~\cite{Coleman:1973jx} 
presented a dynamical mechanism of electroweak symmetry breaking that is based on classical scale invariance. 
Similarly, the framework of technicolor~\cite{Susskind:1982mw} relies on generating scales from strong dynamics in asymptotically free theories such as QCD.
In this work we take this idea one step further and assume that classical scale invariance is a fundamental property of nature.
This principle is similar in essence to the properties of homogeneity and isotropy in cosmology, which are also ultimately broken by
quantum fluctuations.

Classical scale invariance by itself does not solve the hierachy problem, but can be regarded as 
a guideline for model building, which restricts the space of Lagrangians to contain only operators
with dimensionless coupling constants. The hierachy problem then manifests itself 
as the absence of couplings between the Higgs and other energy scales that are dynamically 
generated in the UV. However, it is possible to contemplate that such UV physics is protected by 
symmetries from coupling to the Higgs~\cite{Giudice:2013yca}. 

We illustrate the consequences for model building by constructing a model 
---dark technicolor---in which a new scale is dynamically generated by strong dynamics in a dark 
sector\footnote{A similar construction has been presented in \cite{Hur:2011sv}.}.
We discuss different realisations for creating the dynamical scale $\Lambda_{\rm{TC}}$ in the dark sector: it can arise similarly as in QCD by new physics operating at $\Lambda_{\rm{TC}}$, or by quasiconformal dynamics operating at a scale $\Luv\gg\Lambda_{\rm{TC}}$ leading to an emergent infrared scale $\Lir$ 
as a low energy mirage. This dark scale, then, triggers the EW phase transition via a Higgs portal coupling between the dark sector fields 
and the SM Higgs field. The model explains the existence of cold DM in the Universe with suppressed direct couplings to the SM quarks and leptons.
A dark sector model which leads to EW symmetry breaking via Coleman--Weinberg mechanism has been proposed in Ref.~\cite{Englert:2013gz,Hempfling:1996ht,Chang:2007ki}.

The structure of this paper is the following: In section II we discuss the physical meaning of the naturalness/hierarchy problem and introduce 
the concept of physical naturalness. In section III we discuss the naturalness of the SM and its extensions under approximate scale invariance.
In section IV we present an explicit model example, and we conclude in section V.

\section{Naturalness  - General Considerations}

If naturalness should discriminate between different models in particle physics, then the hierarchy problem (i.e. the stabilisation of the electroweak scale from radiative corrections) must be interpreted as a physical phenomenon. This is what we call ``physical naturalness''. If a UV scale is 
not associated with particles or if those do not couple to the SM in any way, physically measurable contributions to the Higgs mass cannot 
be generated perturbatively.
In particular, the quadratic divergences~\cite{Weinberg:1979bn,Susskind:1978ms,Susskind:1982mw} proportional to
the cut-off scale $\Lambda$ in cutoff regularization schemes are unphysical in the absence of new particles coupled at this 
scale~\cite{Wetterich:1983bi,Bardeen:1995kv,Jackiw:1999qq,Fujikawa:2011zf,Aoki:2012xs,Vieira:2012ex,Englert:2013gz,Farina:2013mla}.
In a single scale theory like the SM Higgs sector, such renormalisation scheme dependent contributions 
should be subtracted from unphysical bare parameters of the Lagrangian together with the infinities appearing in renormalisation~\cite{Jackiw:1999qq,Fujikawa:2011zf,Aoki:2012xs,Vieira:2012ex}. This is manifest in dimensional regularisation, 
and physical observables should be scheme-independent.

\subsection{The standard model}

The SM is a renormalisable quantum field theory with one explicit mass scale given by the Higgs mass term $-\mu^2$ in the scalar potential
\bea
V=-\mu^2 |H|^2 + \frac{\lambda}{4}  |H|^4,
\label{V}
\eea
where $H$ is the Higgs doublet and $\lambda$ is its quartic self-coupling.
All parameters of the SM, including the ones in \rfn{V}, depend logarithmically on the energy scale at which
they are determined. In principle, this dependence is completely determined by renormalisation group (RG) running that introduces no new mass scale 
to the model\footnote{The Standard Model $U(1)$ gauge coupling runs into a Landau pole at $\sim 10^{40}$ GeV. However, this behaviour does not perturbativly generate a quadratic contribution to the Higgs mass. We will briefly comment on the Landau pole in the end of Sec. II B of this paper.}. 

In the limit $\mu^2\to 0$ the SM becomes classically conformal. Classical conformality is broken at loop level by the running of the gauge, Yukawa and Higgs quartic couplings. However, in the SM the RG evolution of $\mu^2$ is proportional to itself, and if one sets as an ultraviolet boundary condition that $\mu^2=0$,
then nonzero $\mu^2$ is not generated in the infrared by the SM couplings\footnote{In fact, $\mu^2\sim\Lambda_{\rm QCD}^2$ will be generated from the Higgs Yukawa couplings to the SM quarks at the QCD confinement scale. However, this process is dynamical and protected from any high scale corrections.}.
Furthermore, due to scale invariance at high energies, there would not be large radiative corrections to the Higgs mass from the new physics. Simply put, since $\mu$ is the only scale in the theory, it follows by dimensional analysis that any radiative correction that has units of mass must be proportional to $\mu$.

By physical naturalness, the SM in isolation is therefore a natural theory in the sense that large radiative corrections to the Higgs mass are absent. While 
being natural, the SM is not a satisfactory theory of fundamental physics as the origin of the EW scale, and the absence of non-renormalisable 
operators remains unexplained.

\subsection{Physics beyond the standard model}

The situation is of course different in models where {\em physical} mass scales do appear at high energies. 
If the SM Higgs couples directly to heavy particles with mass $M$ at that scale, the renormalised Higgs 
mass parameter is proportional to the heavy particle mass and depends logarithmically on the renormalisation scale 
 \bea
m_h^2(\mu) \propto \frac{g^2}{(4\pi)^2} \, M^2 \log\left(  \frac{M^2}{\mu^2} \right),
\label{logcontr}
\eea
where $g$ is the coupling constant that induces the scale mixing and $\mu$ is the renormalisation scale at which the Higgs mass is measured.

The Higgs mass easily receives contributions determined by the high scale alone, even if the new physics operating at high scales is not charged under the SM interactions. This is because the Higgs can couple with scalars $S_i$ via
\bea
\lambda_i |S_i|^2 |H|^2,
\label{portal1}
\eea
independently of the quantum numbers of $S_i$.
If the new scalars get vacuum expectation values (vevs), $\langle S_i \rangle=V_i,$ new contributions to the Higgs mass are generated at tree level. 
If $\langle S_i \rangle=0,$ the SM Higgs mass gets one loop contributions of the order of $\sim \lambda_i/(4\pi)^2 M^2_i$, where $M_i$ is the mass 
of the scalar $S_i$. These contributions have nothing to do with EW symmetry breaking.

A prototypical example is a conventional grand unified theory  (GUT)~\cite{Gildener:1976ai} based on an $SU(5)$ or $SO(10)$ gauge group that is 
broken to the SM gauge group by a GUT-scale Higgs mechanism.
Since the SM Higgs has direct couplings to the heavy gauge and scalar bosons with GUT scale masses,
large contributions of the form \rfn{logcontr} to the SM Higgs mass are generated at one loop.
The naturalness problem in this example is so extensive that supersymmetrising the theory seems to be the most economical way to solve the problem. 
However, the logical possibility remains that there is no GUT and the quantum numbers of the SM particles or any other hints of a GUT model are simply 
coincidences. This possibility might be theoretically unappealing, but we have to note that it is in perfect agreement with the experimental 
non-observation of proton decay. Alternatively, the absence of direct couplings between GUT scale particles and the Standard Model 
can be explained using extra dimensions ~\cite{Hosotani:1983xw,Hosotani:1988bm}.

Another example is provided by right-handed singlet neutrinos $N_i$ that couple to the SM via
\bea
Y_{ij}\bar L^i H N^j,
\label{Y}
\eea
where $L_i$ are lepton doublets and $i,j$ denote generations. If the SM Higgs gets a vev, Dirac masses for the SM neutrinos are generated.
If the the right-handed neutrinos have large Majorana masses $M_i \bar N^c_i N_i$, the smallness of the SM neutrino masses is explained via the seesaw
mechanism~\cite{seesaw} and the baryon asymmetry of the Universe is generated via leptogenesis~\cite{Fukugita:1986hr}.
The large Majorana mass $M$ also generates one loop corrections to the SM Higgs mass,
$\sim Y^2/(4\pi)^2 M^2$, via \Eq{Y}.
Knowing the neutrino mass scale $\sqrt{\Delta m^2_{atm}} \sim 0.05$~eV, and requiring the new mass contribution not to exceed 100~GeV, the
Majorana neutrino mass scale must satisfy $M<10^7$~ GeV.  This scale allows for resonant leptogenesis~\cite{Flanz:1996fb,Pilaftsis:1997jf}
with $M_i\approx M_j.$ Alternatively, the SM neutrinos can be Dirac particles. In this case their lightness requires very small Yukawa couplings $Y$.
In the absence of a theory of flavour, this is a phenomenologically viable scenario as well.

Besides the hypothetical scales associated with the new physics scenarios described above, that may or may not exist, there are two physically defined scales well above the EW scale that are known to exist in some form: the Planck scale and the Landau pole of the $U(1)$ hypercharge of the SM. If these scales are associated with heavy particles or phase transitions, it is likely that they should induce radiative corrections to the low scale physics and thus create the hierarchy problem. In this paper we will work under the assumption that this is not the case.

To justify this assumption, we will shortly discuss both gravity and the Landau pole. First of all, while there are good reasons to believe that classical gravity should be taken over by a quantum theory at or below the Planck scale, there are also reasons to believe that this theory might be very different from the quantum field theories known to us. The Weinberg-Witten theorem \cite{Weinberg:1980kq} states that a massless spin two particle can not exist in a renormalisable Lorentz invariant theory. Also the experimental evidence seems to suggest that gravity is decoupled from particle physics: the smallness of the cosmological constant suggests that gravity does not feel the presence of vacuum expectation values of quantum fields, and the smallness of the Higgs mass suggests that the quantum fields do not feel the presence of the scale of gravity. Thus it is plausible that the UV theory of gravity does not influence low energy physics. As a proof of concept a class of gravity theories has been constructed \cite{Dubovsky:2013ira} in 2D that can not be described by a local Lagrangian, and that do not introduce radiative corrections to the low energy particle physics observables.

Assuming that gravity does not notably influence particle physics, the perturbative description of the SM finally breaks down as the $U(1)$ coupling runs to the Landau pole at $\sim 10^{40}$ GeV. What happens at this scale is not known, but here we will simply assume that the Landau pole signals a smooth transition into a nonperturbative phase of the theory, that does not introduce an explicit violation of scale invariance and does not radiatively affect the low energy observables. Another possible scenario is that the unknown theory of gravity, while decoupled from low energy observables, alters the running of the SM couplings at high energies and that the SM thus avoids the Landau pole. Since the physics above the Planck scale is currently completely unknown, we will not go any deeper in trying to understand what physics may or may not take place in the far UV, but simply assume that whaterver it is, it does not notably influence low energy physics. 

\section{Renormalizability and Classical Scale Invariance}

In this section we will further clarify the role of classical scale invariance in our thinking. It is worth
emphasising again that purely classical scale invariance does not protect IR physics against 
radiative corrections, if scale invariance is broken by quantum effects in the UV. However, 
we will find that it can act as a useful guideline for model building.
We start this section by reviewing the Wilsonian understanding of effective field theories.

\subsection{Renormalizability of effective field theories}

In the Wilsonian approach renormalisability is seen as a consequence of the 
scaling of operators: the Lagrangian is written as a sum of local operators
\be \mathcal{L} = \sum_i \rho_i M_i^{d_i-4} \mathcal{O}_i,
\ee
and defined on a sphere in Euclidean momentum space with radius $M$.  To re-define the theory 
at a lower momentum scale $M'=M/N$, one first integrates out the high-energy modes of the fields in the operators 
$\mathcal{O}_i$ with Euclidean momenta larger than $M/N$. After this, the radius of validity is rescaled back 
to a sphere of radius $M$. This rescaling leads to the new operator coefficients 
\be
\rho_i \to 1/N^{d_i-4} \rho_i . 
\ee
The non-renormalisable operators with mass-dimension $d>4$ have a negative power-law dependence on the renormalization scale, and therefore are naturally absent in the low energy theory. Similarly the super-renormalisable operators with mass-dimension $d<4$, such as the mass term of a scalar field, have a positive power-law dependence on the renormalization scale, and therefore the appearance of such terms with a coefficient that is much smaller than the cut-off scale is seen as unnatural.
\footnote{It should be noted that this (classical) scaling of the operators is merely an effect of rescaling the energy measuring unit. The real physical effects of renormalisation are the logarithmic running of the coupling constants and anomalous dimensions. All explicit mass scales of the effective theory 
are given by the cutoff scale and do not change through renormalization, except for logarithmic 
corrections.}

To summarize, the generic prediction from effective field theories is that:
\begin{enumerate}
\item Relevant operators ($d<4$) have a size equal to the cutoff scale of the theory.
\item Marginal operators ($d=4$) have $\mathcal{O}(1)$ coefficients
\item Irrelevant operators ($d>4$) are suppressed by powers of the cutoff
scale.
\end{enumerate}

Regarding the Standard Model as an effective field theory in the Wilsonian sense then leads to 
the well-known choices: Either the cutoff scale is given by the scale of the only relevant 
operator in the theory (the Higgs mass) at which new physics such as SUSY has to appear. In this case 
we would generically expect e.g.~rapid proton decay from dimension 6 (and higher) 
operators supressed only by the low new physics scale. Avoiding this requires a
conspiracy of nature to hide this cutoff scale from our experimental searches (such as R-parity
to avoid proton decay). 

On the other hand, the cutoff scale could be much larger than the EW scale such that any 
higher order operators are naturally absent. But then it can not be explained why the Higgs 
mass is so much smaller than this scale, and nature would be extremely fine-tuned. 

This dichotomy seems to suggest that the Wilsonian picture can not be 
sucessfully applied to the Standard Model. But how else can we explain the size of the 
Higgs mass and the absence of higher order operators?

\subsection{Classical scale invariance}

Taking the Standard Model at face value, we see that it contains only massless coupling
constants, with the single exception of the Higgs mass parameter. Despite decades of precision
flavour physics, no evidence for any non-renormalisable interactions of Standard Model
fields has been found, and the absence of proton decay suggests that there are no generic 
higher order operators up to the scale of $10^{16}$~GeV. 

But why should nature only allow renormalisable interactions to exist at the fundamental level,
if we disregard the Wilsonian effective field theory explanation? We suggest that the answer may be the underlying classical scale invariance. If we demand that the fundamental theory is scale invariant at the classical level, then (in four dimensions) only operators of mass-dimension $d=4$ are allowed. Any operator of dimension $d\neq4$ must have a dimensionful coupling constant which sets a unique scale in the theory and breaks the classical scale invariance. Thus we will now explore the possibility that classical scale invariance is a fundamental property of nature. 

This property is in essence similar to the invariance of physics of the chosen reference frame or to the Copernican principle in cosmology, which states that no point in space is unique. We now extend these symmetries by demanding that also no momentum scale is unique in the classical 
level. In cosmology, the universe is not isotropic and homogeneous at small scales, because quantum fluctuations have generated structure, and thus 
different points in space have different properties. Similarly the interactions of particles look different at different energy scales, because quantum 
effects break the tree-level scale invariance and generate the running of the coupling constants. Quantum effects can also generate special mass 
scales, such as $\Lambda_{\rm QCD}$, and effective operators of dimension $d\neq4$, but all this is a result of the dynamics of the theory. 
At the classical level the theory is still scale invariant and no momentum scale is set by hand to have a special meaning.

This approach gives us a clear understanding of renormalisability. Fundamental quantum field theories are renormalisable because only renormalisable operators can be embedded in a classically scale invariant theory. It also alters our interpretation of the naturalness problem. The Higgs mass term is unnatural, not because it is small or not protected by symmetries, but because its origin is not explained. If classical level scale invariance is a fundamental property of nature, then this operator can not exist in a fundamental theory. Therefore the SM must be expanded with a mechanism to generate this term dynamically.

\section{An Explicit Model: EW Symmetry Breaking from Dark Technicolor}

As we have discussed, according to physical naturalness the existence of the electroweak scale necessitates the existence of new physics. As a concrete example we take the Higgs field as a fundamental scalar with only quartic interactions. The quadratic term, and hence also the electroweak scale, is generated by dynamical symmetry breaking occurring in the dark sector and transmitted to the visible sector via the Higgs portal term, \rfn{portal1}. As an additional motivation, the model also allows for novel dark matter candidates. 

The most minimal such extension
of the SM is obtained by adding one singlet scalar $S$ to the theory.
Imposing additional $Z_2$~\cite{Silveira:1985rk,McDonald:1993ex,Burgess:2000yq} or
$Z_3$~\cite{Belanger:2012zr} symmetry then allows for scalar DM.
The interpretation of the imposed discrete symmetries is as a matter parity that is a remnant of a gauged
$B-L$~\cite{Kadastik:2009dj,Kadastik:2009cu}.
Adding the singlet to the SM also improves Higgs phenomenology. In the SM the Higgs quartic coupling
runs negative at scale $10^{12}$~GeV~\cite{EliasMiro:2011aa,Xing:2011aa,Kadastik:2011aa} destabilising
the vacuum.
The singlet coupling to the Higgs boson given in \rfn{portal1}, $\lambda_S |S|^2 |H|^2,$ is always allowed independently of the internal
quantum numbers of the singlet, and its contribution to the RG equation of $\lambda$ is always positive
{\it improving} the vacuum stability via RG running. Alternatively, if the singlet gets a vev, the measured Higgs mass
implies larger Higgs quartic coupling than in the SM, {\it improving} the vacuum stability at the tree level\footnote{However, one should note that the question of vacuum stability is more complicated in the presence of several scalar fields than in the SM. In this kind of models the full scalar potential, which has to be bounded from below for vacuum stability, is a function of several couplings. Thus one has to calculate the RG evolution of all the couplings to ensure that the potential is bounded from below at all scales.}~\cite{EliasMiro:2012ay,Lebedev:2012zw}.

However, this simplest scenario fails to explain electroweak symmetry breaking.
Realistic dynamical EW symmetry breaking a la Coleman-Weinberg in the singlet scalar model requires $N\geq12$ singlets with couplings
$\lambda_S>1$ to overcome large top contributions with the wrong sign~\cite{Gildener:1976ih,Hempfling:1996ht,Espinosa:2007qk}. 
If such a strongly coupled $S_i$ is the DM candidate,
this scenario is in conflict with DM direct detection searches~\cite{Kadastik:2011aa}.
A Coleman-Weinberg mechanism via a Higgs portal can be made to work in extended models~\cite{Meissner:2006zh,Kadastik:2009ca,Iso:2012jn,Foot:2007as,Foot:2007iy}.

\subsection{The dark technicolor model}

In this section we work with the minimal singlet extension of the SM, and propose a framework consisting
of two sectors: the visible SM sector respecting classical 
conformal symmetry and the dark sector consisting of SM singlet matter fields, $Q$, interacting via a new strong force, {\em dark technicolor}. 

We start with a scale invariant SM Higgs sector, i.e. the potential for the Higgs field $H$ is
\be
V_H=\frac{\lambda}{4} |H|^4,
\label{Higgspot}
\ee
and $H$ is assumed to be a fundamental scalar field.
We assume that scale breaking originates from the dynamics of a non-trivial dark sector
that manifests itself via the existence of DM. 
The dynamical symmetry breaking in the dark  sector is transmitted to the visible sector via the Higgs portal  \rfn{portal1}.
The relevant part of the Lagrangian is 
\bea
{\cal L} &=& |D_\mu H|^2-\lambda_h(H^\dagger H)^2 +|\partial_\mu S|^2 - \lambda_s|S|^4 -\frac{1}{4}F_{\mu\nu}^a F^{a \mu\nu} \nonumber \\ &+&\sum_{i=1}^{N_f}\bar{Q}_i (i\gamma^\mu D_\mu)Q_i +g_h |S|^2|H|^2-(y_Q S\bar{Q}Q +{\rm{h.c.}}).
\label{dtclagr}
\eea
Here the covariant derivatives operating on the SM Higgs and the dark techniquarks, respectively, contain the SM and dark technicolor gauge fields. The dark technicolor field strength is denoted by $F_{\mu\nu}$, and we omitted the SM contributions beyond the Higgs sector as well as the Yukawa couplings of the Higgs to SM matter fields.
A similar setup of two sectors was used in \cite{Englert:2013gz,Hempfling:1996ht,Chang:2007ki}  to generate dynamical EW symmetry breaking from a dark Coleman-Weinberg mechanism. A model that is similar to our proposal was presented in~\cite{Hur:2011sv}, based on two flavor QCD-like theory in the hidden sector. In the following we will investigate a broader range of possibilities for the dark technicolor dynamics.

We assume that at strong coupling, spontaneous chiral symmetry breaking happens in the dark technicolor sector analogously to QCD, and leads to the 
dynamical generation of a scale $\Lambda_{\rm TC}$. Depending on the fermion representation, the possible global symmetry breaking patterns are
\bea
&&{\rm{SU}}(N_f)\times{\rm{SU}}(N_f)\to {\rm{SU}}(N_f),\,\,{\rm{(complex)}}\nonumber \\
&&{\rm{SU}}(2N_f)\to {\rm{Sp}}(2N_f),\qquad \qquad\,{\rm{(pseudoreal)}} \nonumber \\
&&{\rm{SU}}(2N_f)\to {\rm{SO}}(2N_f),\qquad \qquad {\rm{(real)}} 
\eea

This has important consequences for the dark matter candidates: The Goldstone bosons associated with the above symmetry breaking pattern will become pseudo-Goldstones due to the assumed coupling with the Higgs portal. In the case of complex representations, all of these approximate Goldstones are $\bar{Q}Q$- states, and due to the hidden sector flavor symmetry the lightest Goldstone is stable and becomes a dark matter candidate. The new matter fields allow for a U(1)$_{\rm{dark}}$ symmetry, analogous to ordinary baryon number, which will protect the lightest state carrying a nonzero charge under this symmetry.
For complex representations those states are heavy in comparison to the pseudo-Goldstone bosons.
An appealing feature of (pseudo)real representations is that the pseudo-Goldstones are $QQ$-states and can carry also nonzero U(1)$_{\rm{dark}}$ charge. Hence, some of the Goldstone bosons of the spontaneously broken global symmetry are also dark baryons. The lightest of these is a natural dark matter candidate.

As a concrete example we take $N_f=2$ in the adjoint representation\footnote{As explained above, this is different from the model investigated in \cite{Hur:2011sv}, where the dark technifermions transform under a complex representation. Our choice allows for a light pseudo-Goldstone boson carrying technibaryon number, whereas the technibaryons are heavy in the case of complex representation.}. Then the global symmetry 
breaking pattern is SU(4)$\rightarrow$ SO(4) and the number of Goldstones is nine.
If the number of dark colours is sufficiently large, then the theory is QCD-like, i.e. far
below the conformal window e.g. $N>4$. \cite{Sannino:2004qp,Dietrich:2005jn}. In this case, we can 
apply our established intuition about the features of the dark sector. In particular, it also imposes that the associated energy scale 
$\Lambda_{\rm{TC}}$ cannot be too far above the EW scale if it is to source the EW scale of the SM. Consequently, 
there are no large radiative corrections to $\Lambda_{\rm{TC}}$ and hence, no large radiative corrections to the EW scale.

However, there is no need for the dark sector to be QCD like. If we suppose that the underlying theory is $N_f=2$ fermions in the adjoint representation 
of $SU(2)$ or $SU(3)$, then there is growing evidence from lattice simulations that this theory has an infrared fixed point, see {\it e.g.} \cite{Catterall:2007yx,Hietanen:2009az,Hietanen:2008mr,DelDebbio:2010hx,DeGrand:2011qd}. Note that these results apply only to strongly interacting theory in isolation. In our case the dark techniquarks have a Yukawa coupling with the singlet field $S$. The standard analysis of Schwinger-Dyson equation in the ladder approximation \cite{Yamawaki:1996vr} shows that
in the presence of nonzero Yukawa coupling $y_Q$ the critical coupling for the formation of dark techniquark condensate, $\alpha_c$, decreases with respect to its value at $y_Q=0$. This in turn implies that the theory which at $y_Q=0$ is observed (say, on the lattice) to be inside the conformal window may break chiral symmetry and hence move outside the conformal window when nonzero Yukawa coupling is turned on. Such theory, under the RG evolution from the UV towards IR feels the presence of the fixed point, but is ultimately repelled away from it leading to the development of an infrared scale $\Lir$. This situation of quasi-conformality is known as walking. In light of our model, it implies that the dynamical interactions of the theory occur at very high scales, $\Lambda_{\rm{UV}}$, where the theory flows towards the fixed point. 

In relation to our earlier notation $\Lambda_{\rm TC}=\Lir$. The low energy phenomenology, coupling to the portal scalar and through that to the SM Higgs, proceeds as in the case considered earlier. The essential difference is that now approximate scale invariance---walking---guarantees that there are no large radiative corrections to $\Lambda_{\rm TC}$ even if the new physics operates at very high scales, $\Luv\gg\Lir$. Consequently, there are no large corrections to the EW scale.  

Therefore, we have two possibilities underlying the dark sector: 
First, all new physics occurs at the scale 
$\Lambda_{\rm{TC}}\sim\Lir$ such that $\Lir/\vw\sim 10\dots 10^4$. Second, new physics is governed by a quasi stable fixed point at scale $\Luv \gg \vw$. However, the RG evolution is driven away from the vicinity of this fixed point by the formation of the chiral condensate above the scale $\Luv$, but leading to a dynamically generated infrared scale $\Lir\sim \Lambda_{\rm{TC}}$. These two possibilities realize the general new physics scenarios we have discussed in previous sections.

\subsection{Low energy phenomenology}

In both cases considered above, the low energy dynamics of the strongly coupled dark sector is most conveniently treated in terms of the low energy degrees of freedom, i.e. the Goldstones. Using a linear representation, the potential of the low energy effective theory for the dark sector is
\be
V=-\frac{m^2}{2}{\rm{Tr}}(M^\dagger M)+\lambda_1{\rm{Tr}}(M^\dagger M)^2+\frac{\lambda_2}{4}({\rm{Tr}}(M^\dagger M))^2,
\ee
where $M$ is the dark meson matrix. Explicitly, we write\footnote{For explicit realization of the SU(4) generators and the matrices $X^a$, see \cite{Foadi:2007ue}.}
\be
M= \frac{1}{2}\left(\sigma+i\eta+2\sqrt{2}(\sigma^a+i\pi^a)X^a\right)E
\ee
where $\sigma$ is a scalar field component that gets a vacuum expectation value 
$\langle \sigma \rangle =v_\sigma$ and $X^a$ are the broken generators of the symmetry breaking
pattern $SU(4)\rightarrow SO(4)$. The fields $\sigma^a$ are the scalar partners of the pseudoscalars $\pi^a$, and their presence is required to realize the global symmetry linearly and to guarantee the transformation $M\rightarrow gMg^T$ with $g\in$SU(4). The matrix $E$ in this case is
\be
E=\begin{pmatrix} 0 & 1 \\ 1 & 0 \end{pmatrix}.
\ee

First, let us review how electroweak symmetry breaking arises in this model setting.
For simplicity, we assume that the dark sector is connected to the visible sector via a singlet scalar $S$. As we have already discussed, this scalar couples to the Higgs, $|S|^2|H|^2$, while gauge invariance forbids any other renormalizable coupling. The Yukawa coupling between the scalar and the dark techniquarks also 
induces a direct coupling between the messenger field $S$ and the $\bar{Q}Q$ condensate; $S\langle \bar{Q}Q \rangle \sim \Lambda_{\rm TC}^3 S$. 
The vacuum expectation values of the scalar field $s$ and the scalar components $h$ and $\sigma$ of $H$ and $M$ are determined from the potential 
\be
V_s=\lambda_h h^4+\lambda_\sigma (\sigma^2-v_\sigma^2)^2-g_h s^2h^2+g_\sigma s^2\sigma^2+\lambda_s s^4+v_\sigma^3 s,
\label{EffScalarPotential}
\ee
where $\lambda_\sigma=(\lambda_1+\lambda_2)/4$. Replacing first $\sigma$ with its vev $v_\sigma\sim\Lambda_{\rm TC}$, the extremum conditions imply
\bea
\langle s\rangle &=& \alpha v_\sigma,\nonumber\\
\langle h\rangle &=& \frac{\sqrt{g_h}}{\sqrt{2\lambda_h}}\langle s \rangle,
\label{shvev}
\eea
where
\be
\alpha=\frac{2\sqrt[3]{3}g_\sigma\lambda_s+\left(9\lambda_s^2-\lambda_s\sqrt{3\lambda_s(8g_\sigma^3+27\lambda_s)}\right)^{\frac23}}{2\sqrt[3]{9}\lambda_s\left(9\lambda_s^2-\lambda_s\sqrt{3\lambda_s(8g_\sigma^3+27\lambda_s)}\right)^{\frac13}}.
\label{equationAlpha}
\ee
The vev of the messenger field $s$ is suppressed with respect to the dark technicolor scale by a factor of 
$\alpha\sim g_\sigma$, where $g_\sigma$ is the effective coupling between the messenger and the composite fields induced by the Yukawa coupling $S\bar{Q}Q$. The vev of the Higgs field is suppressed by an additional factor of $\sqrt{g_h}/\sqrt{\lambda_h}$, where $g_h$ is the portal coupling constant which we assume to be small. Expanding in the small parameter $g_h$ the masses of the scalar states read
\bea
m_\sigma^2&=&(8\lambda_\sigma+2g_\sigma\alpha)v_\sigma^2+\mathcal{O}(g_h),\nonumber\\
m_s^2&=&4g_\sigma\alpha v_\sigma^2+\mathcal{O}(g_h),\nonumber\\
m_h^2&=&4g_h\alpha^2 v_\sigma^2+\mathcal{O}(g_h^2).
\label{masses}
\eea
Here the subscript on the masses refers to the dominant component of the mass eigenstate, so that $m_\sigma$ is the mass of the scalar state that, after diagonalising the mass matrix, has the largest component of the composite field $\sigma$ etc. The mixing pattern also follows the same hierarchical structure as the masses and vacuum expectation values of the fields; the mixing angle between $\sigma$ and $s$ is of the order of $\mathcal{O}(g_\sigma^2)$, and the mixing between $h$ and $s$ is $\mathcal{O}(g_h^{3/2}g_\sigma)$. There is no direct mixing between $\sigma$ and $h$. 

In addition to generating the electroweak scale from dynamical symmetry breaking in the dark sector, the model also provides DM candidates. As we have discussed, the properties of this WIMP are dependent on the fermion representation---and the associated symmetry breaking pattern---in the dark sector. In our case of $N_f=2$ dark techniquarks in the adjoint representation, three of the nine Goldstones are neutral under the baryon charge and the remaining six are baryons.
 
The dark technibaryon is produced in thermal equilibrium with the SM in the early universe, and the DM density then freezes out as usual for a thermal relic. The main annihilation channel of the (lightest) dark technipion $\pi_0$ is through the portal into SM Higgs bosons. With the notation introduced above, this implies the cross section
\be
\langle v\sigma \rangle \sim \frac{g_\sigma^2g_h^2\alpha^4v_\sigma^4}{512\pi m_\pi^6}\frac{\sqrt{1-\frac{m_h^2}{m_\pi^2}}}{(1-\frac{m_s^2}{4m_\pi^2})^2}.
\ee
This will be modified by a small overall factor due to the mixing of the scalars, but this effect will be compensated for by the fact that due to mixing, dark technipions can also annihilate to SM particles via the direct coupling with $h$. To get a simple estimate, we use  (\ref{masses}) to express $v_\sigma$ in 
terms of $m_s$, and obtain
\be
\langle v\sigma \rangle\sim \frac{g_h^2\alpha^2}{8192\pi}\frac{\hat{m}_s^4\sqrt{1-\hat{m}_h^2}}{(1-\frac{1}{4}\hat{m}_s^2)^2}\frac{1}{m_\pi^2},
\label{annihilation_xsec}
\ee
where we also defined $\hat{m}_i\equiv m_i/m_\pi$.
Hence, the relevant parameters for the dark matter relic density are the portal coupling $g_h$, the coupling $\alpha$ defined in (\ref{equationAlpha}), the mass of the dark technipion $m_\pi$ and its relative magnitude with respect to $m_s$ and $m_h$. As we have already discussed, the natural ordering of the masses in the scalar sector is
\be
m_\sigma^2\gg m_s^2 \gg m_h^2.
\ee
The pion mass originates from the explicit chiral symmetry breaking term $y_Q S\bar{Q}Q$ in (\ref{dtclagr}), and is suppressed by the Yukawa coupling $y_Q$ of the dark techniquarks and by the coupling $g_\sigma$ from the scale set by $v_\sigma$. Hence, we clearly have $m_\sigma^2\gg m_\pi^2$. 
Then, depending on the relative magnitude of $y_Q$ and $g_\sigma$, the phenomenologically interesting regions are $m_s\gg m_\pi>m_h$ or $m_s\gtrsim m_\pi>m_h$.

\begin{figure}[t]
\begin{center}
\includegraphics[width=0.49\textwidth]{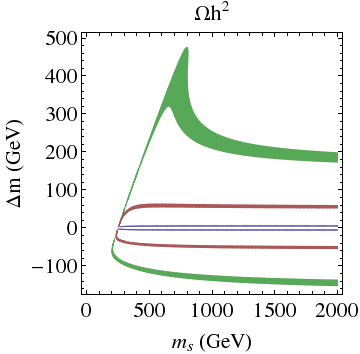}
\caption{The $5\sigma$-allowed region for the relic density of DM, for $g_h\alpha=0.3$ (green), $g_h\alpha=0.1$ (red) and $g_h\alpha=0.01$ (blue). The mass of the messenger scalar is plotted on the $x$-axis, and the $y$-axis shows $\Delta m=m_s-2m_\pi$.}
\label{fig:relicdensity}
\end{center}
\end{figure}

The Planck Collaboration \cite{Ade:2013lta} measured the cold DM relic
density to be $\Omega_c h^2 \pm \sigma = 0.1199 \pm 0.0027$.
From the annihilation cross section (\ref{annihilation_xsec}), we can estimate \cite{Jungman:1995df}
\be
 \Omega h^2 \simeq \frac{ 3 \times 10^{-27} {\rm cm}^3 \, {\rm s}^{-1}
}{\langle v\sigma \rangle}.
\ee
 We plot the experimentally allowed region, at $5\sigma$ confidence level, for the thermal relic abundance of the DM in figure \ref{fig:relicdensity}, for the choise of parameters $g_h\alpha=0.3$ (green region), $g_h\alpha=0.1$ (red region) and $g_h\alpha=0.01$ (blue region). The mass of the messenger scalar is plotted on the $x$-axis, and the $y$-axis shows the deviation from the resonant anihilation region $\Delta m=m_s-2m_\pi$. If the portal coupling $g_h$ is very small, the correct relic density is reached only very close to the resonant region. If the coupling is larger, the allowed region is wider, but a large coupling would imply measurable deviations from the SM in the Higgs couplings due to the mixing with $s$, which in this case would not be very much suppressed. We will return to this point in a future work.

Of course, having a dark sector offers a wider range of possibilities than a single WIMP candidate: there can be several different particles contributing to the observed dark matter abundance, and their relic densities can be either of thermal origin or created by an asymmetry similar to the ordinary baryon-antibaryon asymmetry. As explained above, there are two possible symmetries that can make dark matter candidates stable. First, there is a dark baryon number resulting from the global $U(1)_{\rm dark}$ symmetry in the dark sector, which will protect the stability of the lightest state that carries baryon number. In case of a complex representation these states are heavy, of the order of the confinement scale, but in case of a (pseudo)real representation the lightest baryons are Goldstone bosons and hence light. This was the scenario investigated above. Second, if the dark sector contains several non-degenerate flavors, there will be an unbroken flavor symmetry protecting the lightest non flavor singlet state. In case of a complex representation this is the lightest stable particle, since the baryons are heavy. In case of a (pseudo)real representation, this state and the lightest baryon are both possible dark matter candidates and may both contribute to the dark matter abundance. A detailed investigation of the dark matter abundance and collider phenomenology of this model will be carried out in a future work.

\section{Discussion and Conclusions}

Over past decades the prevailing implication of naturalness for model building has been the cancellation of the quadratic sensitivity of the SM Higgs mass to UV physics. 
At the same time the LHC, flavor physics and electroweak precision results all indicate the absence of any new physics, which is however required  
by the observation of DM. The LHC Higgs data alone suggests that the SM is a consistent theory up to at least $10^{12}$~GeV. 

In the second part of this paper we have reviewed the hierachy problem as a problem of the coupling of scales. If any UV physics scale exists and couples to 
the Higgs, we would expect the Higgs mass to acquire corrections of the order of
this UV mass scale. We have here assumed that no such UV mass scale exists. We then 
explored the consequences of this speculative assumption.

As a guideline for model building we have suggested classical scale invariance, 
which implies that all mass scales are ultimately generated by quantum effects. 
We have pointed out that classical scale invariance as a principle is analogous to the invariance of physics from the choice of a particular reference frame, or to the Copernican principle of cosmology that there are no preferred points in space.
Consequently, any distinct scale of physics must then be generated by dynamical effects. 

We have studied a scenario in which new physics exists at the TeV scale, not to stabilise the EW scale but to create it.
Taking the existence of DM as an additional guiding principle, we proposed the existence of two sectors, the SM sector which is classically 
scale invariant, and the dark sector that we assumed to be a strongly coupled gauge theory. 
They are connected by the Higgs portal. We have presented an explicit model generating the negative SM Higgs mass term from 
dark sector singlets. The model can accommodate the observed amount of DM particles, which we identify with the lightest dark baryon.
This theory can be tested via the Higgs portal coupling to the dark sector, which is predicted to be small.

The dark technicolor model explains the origin of the electroweak scale and dark matter. However, open questions remain, such as the origin of the flavor structure and mass hierarchy of the SM fermions, that likely necessitate the existence of additional fields and dynamics. According to the paradigm of physical naturalness, any new particles related to such models must either be reasonably light, or there has to be a mechanism that protects the Higgs mass and the electroweak scale from radiative corrections due to couplings with these particles. Our proposal of physical naturalness suggests to search for a solution 
by embedding this new model as a whole in a classically conformal UV-theory.

\mysection{Acknowledgements}
We thank A. Strumia for discussions.
This work was supported by the ESF grants 8499, 8943, MTT8, MTT59, MTT60, MJD140, MJD435, MJD298, 
by the recurrent financing SF0690030s09 project and by the European Union through the European Regional Development Fund.

\end{document}